\documentclass[a4paper,12pt,twosided]{article}
\usepackage{fancyhdr}
\usepackage{caption}
\usepackage{amssymb,amsmath,amsthm,graphics}
\usepackage[driver]{graphicx}

\begin{document}

\lhead{\sf{P. Filip}}
\chead{}
\rhead{\bfseries \sf{Elliptic Flow in Collisions of Deformed Nuclei
}}
\vspace{2cm}
\begin{center}
\sf\textbf{\Large{Nuclear geometry and elliptic flow in\\
collisions of deformed nuclei}}\\
\vspace{1cm}
\sf\textbf{\underline{P. Filip}\\}
\vspace{0.5cm}
\sf\em{Institute of Physics, Slovak Academy of Sciences, Bratislava
}
\end{center}

\section{Introduction}
\indent

Elliptic flow measurement \cite{elliptMeasure,elliptCumul} has become a very 
important method for determining properties of dense strongly interacting matter
created in ultra-relativistic heavy ion collisions experiments.
Being theoretically predicted in 1992 \cite{JYO1992}
the elliptic flow has been observed in heavy ion collisions
at AGS, SPS and RHIC \cite{EllipticRHIC}.

The elliptic flow is generated during the expansion 
of compressed QCD matter into vacuum from the initial 
asymmetrical volume created in non-central collisions
of nuclei.
During the process of expansion 
(which can be described by hydrodynamical \cite{JYO1992}
or rescattering \cite{Humanic} models) an asymmetry in the azimuthal
distribution of transverse momenta of particles is generated.

Experimentally measured azimuthal distribution of 
particles in non-central heavy-ion collisions  
exhibits second-order oscillation
of the form $\rho(\phi) = a[1+2 v_2\cdot \cos(2\phi)]$,
where parameter $v_2$ characterizes the strength of the 
elliptic flow effect \cite{elliptMeasure}.
Quantitative comparisons of the initial spatial eccentricity $\varepsilon$
and final elliptic flow strength $v_2$ allow one to make implications on the
equation of state of strongly-interacting QCD matter \cite{EOSv2}.

In this contribution we investigate the elliptic flow behaviour
in relativistic collisions of deformed nuclei.
Using Optical Glauber Model simulation \cite{MyOGM}
we predict large fluctuations of the elliptic flow 
(at given collision centrality) in collisions of deformed nuclei.
We also suggest that behaviour of the elliptic flow in the most central
collisions of deformed nuclei is different for nuclei with 
oblate and prolate deformations.

\section{Elliptic flow in collision of spherical nuclei}
\indent

Initial eccentricity $\varepsilon $ of the compressed QCD matter
in collisions of spherical nuclei (e.g. Pb$^{207}$) is simply related
to the size of impact parameter $ b $ (a distance
between centers of colliding nuclei in transversal plane).
Collisions of spherical Pb nuclei with fixed impact parameter 
(e.g. $b = 3\,$fm) will exhibit elliptic flow values fluctuating 
around some average value $\langle v_2 \rangle$. 
Fluctuations of $v_2$ values in such collisions with 
fixed impact parameter originate e.g. 
from the fluctuations of initial participant eccentricity which
occur due to varying positions of individual nucleons \cite{NNfluct}
inside the interaction volume. Thus even in the most central
collisions of spherical nuclei ($b = 0\,$fm) the eccentricity 
$\varepsilon $ fluctuates and on average $\langle \varepsilon \rangle > 0$. 

These effects can be studied in Monte Carlo Glauber Model simulations \cite{MCG}
where individual positions of
nucleons are determined for each collision event. In the optical Glauber model
such effects are neglected.

In the next sections we predict event-by-event fluctuations of the 
elliptic flow $v_2$ due to initial eccentricity $\varepsilon $ fluctuations 
originating from the deformed shape of colliding nuclei. 
Such $v_2$ fluctuations (present only in collisions of deformed nuclei) 
are generated additionally, on top of $v_2$ fluctuations 
observed in collisions of spherical nuclei.

\section{Elliptic flow in collisions of deformed nuclei}
\indent

In collisions of deformed (oblate or prolate) nuclei the initial
eccentricity $\varepsilon $ of the interaction zone strongly depends on the
orientation of nuclear spin (axis of the ellipsoid). For example,
in central ($b = 0\,$fm) collision of two prolate nuclei (e.g. Ho$^{165}$)
the eccentricity can be large if spins of Ho$^{165}$ nuclei 
are parallel to each other and orthogonal to the beam axis,
or negligibly small for spins of both Ho$^{165}$
nuclei being parallel to the beam axis. Eccentricity $\varepsilon $ thus depends on 
azimuthal and polar angles $\phi$, $\theta $ of nuclei colliding: 
\begin{equation}
\varepsilon [b] = f (\theta_1, \phi_1, \theta_2,\phi_2)
\end{equation}

Using the optical Glauber simulation \cite{MyOGM} one can calculate 
number of participating nucleons $N_{\text{part}}$, number of nucleon-nucleon
collisions $N_{\text{coll}}$ and eccentricity $\varepsilon \,$ for
any given value of impact parameter $b$ and angles 
$\phi_1,\theta_1,\phi_2,\theta_2$ (see Fig.1).

\begin{figure}[h]
\begin{center}
\includegraphics[width=8.5cm]{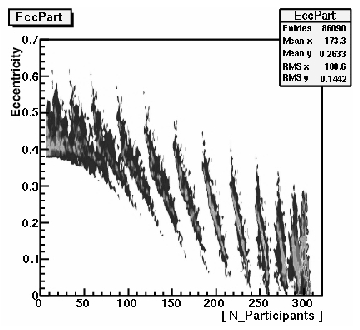}
\includegraphics[width=3.8cm]{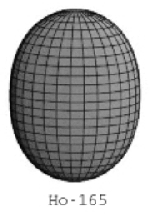}
\setlength{\captionmargin}{66pt}
\caption{\label{fig:FF1}
{\small 
Contour-plot of [ $\varepsilon$ ; $N_{\text{part}}$] values (left) obtained
with optical Glauber simulation of Ho$^{165}$+Ho$^{165}$ collisions
using fixed impact parameters $b=0,1,2,\ldots 12\,$fm,
and the shape of Ho$^{165}_{7/2-}$ nucleus with $\beta_2=0.3$ (right).
} }
\end{center}
\end{figure}

\vspace{-0.9cm}

\subsection{Prolate nuclei}
\indent

Let us consider rare-earth nucleus Ho$^{165}$. This element has a single
stable isotope with prolate ground-state deformation $\beta_2 = 0.3$ 
\cite{Moller}. Using deformed Woods-Saxon density \cite{PRC_defWS}
the eccentricity of the interaction zone
\begin{equation}
\varepsilon = \frac{\sqrt{(\sigma_y^2 - \sigma_x^2)^2 + 4\sigma_{xy}^2}}{
                    \sigma_y^2+\sigma_x^2}
\end{equation}
(here $\sigma_{xy} = \langle x\cdot y\rangle - \langle x\rangle \langle y \rangle$)
can be evaluated (e.g. using $N_{\text{coll}}[x,y]$ density in transversal 
plane) for all combinations of independent angles 
$\theta_1$,$\theta_2 \in\langle 0,90\rangle $ and 
$\phi_1, \phi_2 \in \langle 0,180)$. This has been done 
in 15 degree steps for impact parameters $b = 0,1,2,\ldots 14\,$ fm 
(see Fig.1). 
Every given collision event with 
angles $\theta_1 , \theta_2$ has been weighted by spherical angle factor
$\sin(\theta_1)\sin(\theta_2)$ and impact parameter factor\footnote{
To obtain non-zero contribution from $b=0\,$fm collisions factor $2\pi(b+0.2)$fm
has been used.} $2\pi \cdot b\,$
to account for the probability of the collisions as they occur in
unpolarized experiments. 
In the resulting histogram of ($\varepsilon; N_{\text{part}}$)
values (as shown in Fig.1)
one observes large fluctuations of eccentricity $\varepsilon$ 
for collisions with fixed impact parameter $b$. 
Also number of participants varies at given
impact parameter value due to variation of angles $\theta_1,\theta_2 $ 
and $\phi_1,\phi_2 $.

\subsection{Oblate nuclei}
\indent

Almost all simulations of relativistic Au+Au collisions assume the gold nucleus
Au$^{197}_{3/2+}$ to be spherical. However, this nucleus 
with quadrupole moment $Q\approx 0.55\cdot 10^{-24}$ cm$^2$ \cite{Pykko}
is predicted to have slight oblate 
deformation in the ground state with $\beta_2 \approx -0.13 $ \cite{Moller}.
In order to investigate possible consequences of the predicted 
oblate deformation of Au$^{197}$ nucleus in heavy ion collisions 
we assume it to be deformed with $\beta_2 = -0.131 $ \cite{Moller}.
The result of our optical 
Glauber model simulation is shown in Fig.2. One observes fluctuations of
eccentricity and number of participants at given fixed impact parameters
$b = 0, 1, 2, \ldots \,$fm. For spherical nuclei the optical Glauber
simulation would give a single value of the eccentricity and the number of participants
for any given impact parameter $b$.

\begin{figure}[h]
\begin{center}
\includegraphics[width=15.5cm]{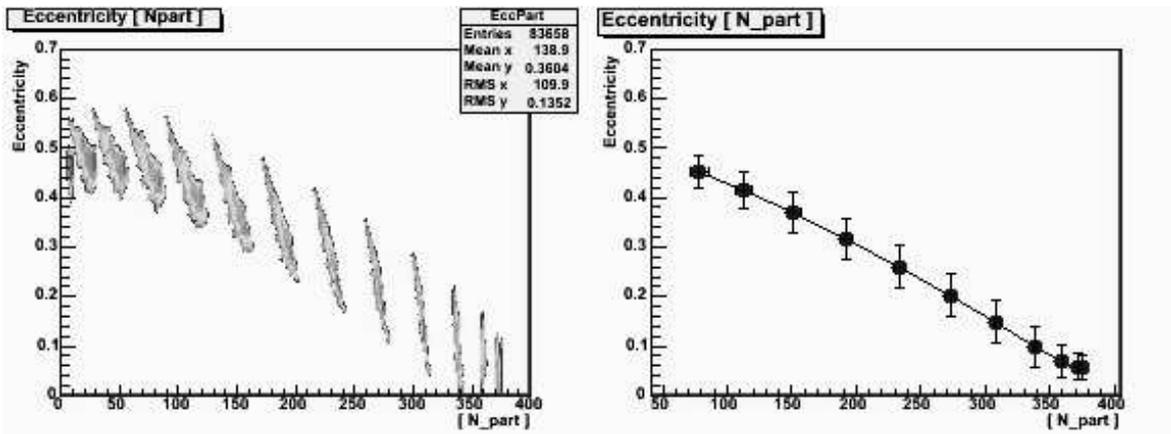}
\setlength{\captionmargin}{33pt}
\caption{\label{fig:FF2}
{\small
Contour-plot (left) of [ $\varepsilon$ ; $N_{\text{part}}$] values obtained
for Au$^{197}$+Au$^{197}$ collisions
at $b=0,1,2,\ldots 12\,$fm assuming $\beta_2 = -0.13$. Error bars in the 
Ecc[$N_{\text{part}}$] plot (right) show the width of eccentricity fluctuations
at given fixed impact parameter values.
} }
\end{center}
\end{figure}

\vspace{-0.3cm}

One can conclude that ground-state deformation of nuclei used in
heavy-ion collision experiments generates additional 
initial eccentricity (and consequently $v_2$) fluctuations 
on top of those existing in the collisions of spherical nuclei.

\section{Central collisions of deformed nuclei}
\indent

In the previous section we have shown that oblate and prolate
deformations of nuclei generate significant fluctuations of 
the initial eccentricity at given impact parameter of collisions.

What happens in the most central collisions ? To investigate this
one has to calculate properly the centrality of collisions using
number of participants $N_{\text{part}}$ and number of nucleon-nucleon
collisions $N_{\text{coll}}$ obtained from the optical Glauber simulation.
We use two-component model \cite{Kharzeev} to calculate
charged-particle multiplicity 
in the form:

\begin{equation}
dN_{\text{ch}}/d\eta =
   (1-x)\cdot n_{pp}\frac{N_{\text{part}}}{2} +x\cdot n_{pp}\,N_{\text{coll}}
\end{equation}
and quantity $dN_{\text{ch}}/d\eta$ is assumed to be the collision centrality
measured experimentally. Results shown in Figure 3 have been obtained
using $n_{pp}=2.25$ and $x=0.11$ \cite{Kharzeev}. Since the number 
$N_{\text{coll}}$ is significantly larger compared to $N_{\text{part}}$, 
approximately 40\% of
secondary particles originate from quantity $N_{\text{coll}}$ and 
the rest of particle multiplicity is generated
due to $N_{\text{part}}$. 

\begin{figure}[h]
\begin{center}
\includegraphics[width=15.5cm]{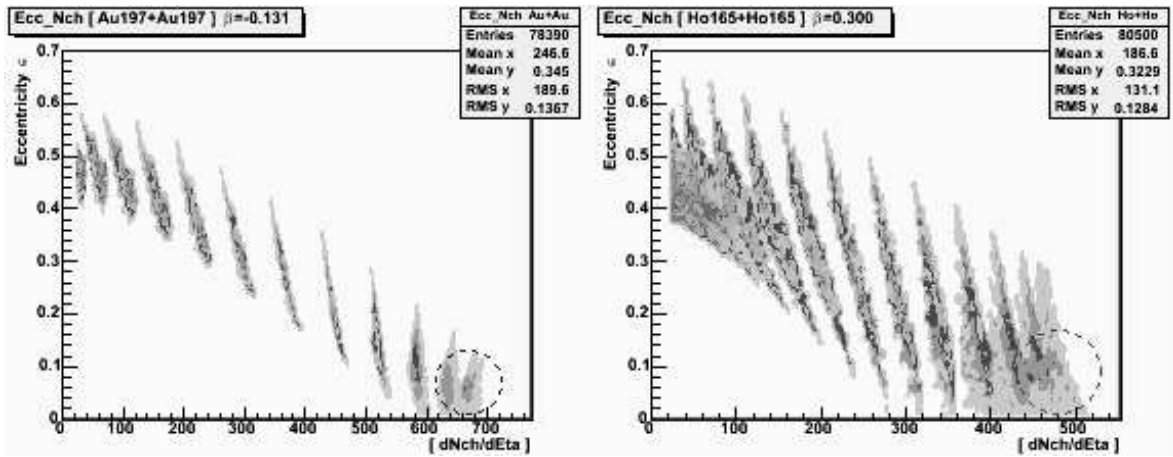}
\setlength{\captionmargin}{33pt}
\caption{\label{fig:FF3}
{\small
Contour-plots of [ $\varepsilon$ ; $dN_{\text{ch}}/d\eta$ ] values obtained
with optical Glauber simulation 
for Au$^{197}$+Au$^{197}$ collisions (left) and Ho$^{165}$+Ho$^{165}$ collisions 
(right) at $b=0,1,2,\ldots 12\,$fm. 
} }
\end{center}
\end{figure}

\vspace{-0.3cm}
 
One observes (see also Figure 4) that most central collisions of oblate 
and prolate nuclei exhibit different behaviour. 
For prolate nuclei the highest multiplicity
of secondary particles is predicted to happen when $b=0\,$fm and spins of 
Ho$^{165}$ nuclei are parallel to the beam axis and to each other. In this case
the number of nucleon-nucleon collisions is significantly higher compared
to the case when spins of Ho$^{165}$ nuclei colliding at $b=0\,$fm 
are orthogonal to the beams axis (and parallel to each other). Therefore
we predict the elliptic flow values at very-high-multiplicity (VHM)
collisions of prolate nuclei to decrease significantly (see Fig.4).

For oblate nuclei our optical Glauber model \cite{MyOGM} predicts 
the opposite effect.
The elliptic flow at most central (VHM) collisions should stay non-zero,
and even slightly rise up ! This small increase of the eccentricity
values in the highest multiplicity Au+Au collisions can be understood easily:
For central collisions ($b=0\,$fm) the highest number of nucleon-nucleon
collisions $N_{\text{coll}}$ is obtained when spins of two oblate nuclei are
parallel to each other and orthogonal to the beam axis. In this configuration
eccentricity $\varepsilon$ of the interacting volume is non-zero 
(being proportional to deformation parameter $| \beta_2|$ of the oblate nuclei).
Increasing the relative {\it azimuthal\,} angle of nuclear spins from 
$0\rightarrow 90$ degrees in this configuration causes the eccentricity 
to vanish and  number of $n-n$ collisions $N_{\text{coll}}$ slightly decreases.
 
\begin{figure}[h]
\begin{center}
\includegraphics[width=15.5cm]{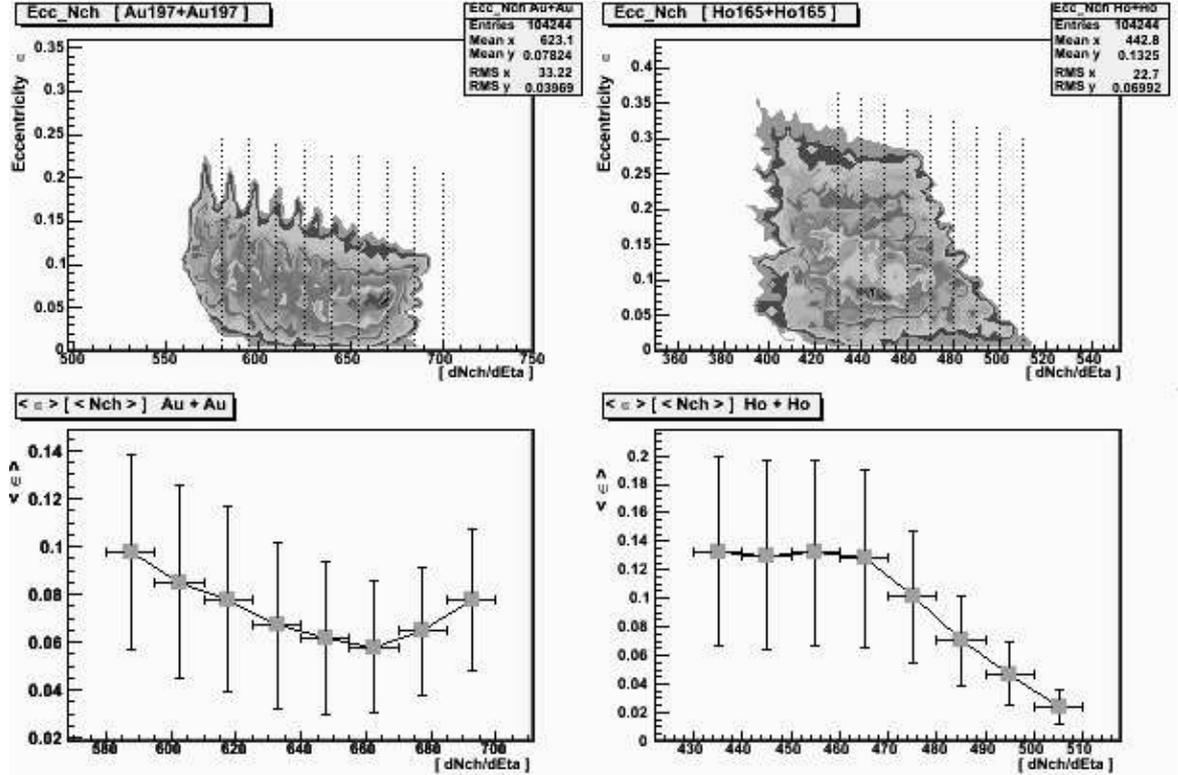}
\setlength{\captionmargin}{22pt}
\caption{\label{fig:FF4}
{\small
Contour-plots of [ $\varepsilon$ ; $dN_{\text{ch}}/d\eta$ ] values obtained
for central Au$^{197}$+Au$^{197}$ (left) and Ho$^{165}$+Ho$^{165}$ 
collisions (right) at $b=0.0,0.2,\ldots 3.2\,$fm. Error bars in bottom
plots show the width of eccentricity fluctuations in the regions
of $dN_{\text{ch}}/d\eta$ indicated in contour-plots (top).
} }
\end{center}
\end{figure}

\vspace{-0.3cm}

This explains results shown in Fig.4 obtained for very high multiplicity Au+Au
collisions simulated with impact parameters $b=0.0, 0.2, \ldots 3.2\,$fm.
One should keep in mind that eccentricity values and fluctuations obtained
by our optical Glauber model are subject to additional fluctuations 
originating from individual positions of interacting nucleons \cite{NNfluct}
which are taken into account only in Monte-Carlo Glauber model simulations 
\cite{MCG}. Therefore,
effects predicted here for Au+Au collisions may require high statistics
and precise elliptic flow measurements at RHIC. 

\section{Comparison with RHIC data}
\indent

In heavy ion collisions experiments one can measure only the elliptic 
flow strength $v_2$ (not the initial eccentricity $\varepsilon\,$). However, 
hydrodynamical scenario \cite{JYO1992} for the expansion of QCD matter 
created in these collisions predicts the elliptic flow to scale 
with initial eccentricity $v_2 \approx \kappa\cdot \varepsilon$. 
This allows one to compare relative strength of $v_2$ fluctuations 
$\sigma_{v_2}/\langle{v_2}\rangle$
with the relative strength of eccentricity fluctuations
$\sigma_{\varepsilon}/\langle{\varepsilon}\rangle$. 

In Figure 5 we compare results of Glauber Monte Carlo simulation 
(assuming Au$^{197}$ nucleus to be spherical) and experimental data obtained by 
PHOBOS collaboration \cite{PHOBOS} with
results of our optical Glauber simulation assuming Au$^{197}$ nucleus
to be slightly deformed ($\beta_2 = -0.131$). 

\begin{figure}[h]
\begin{center}
\includegraphics[width=8.8cm]{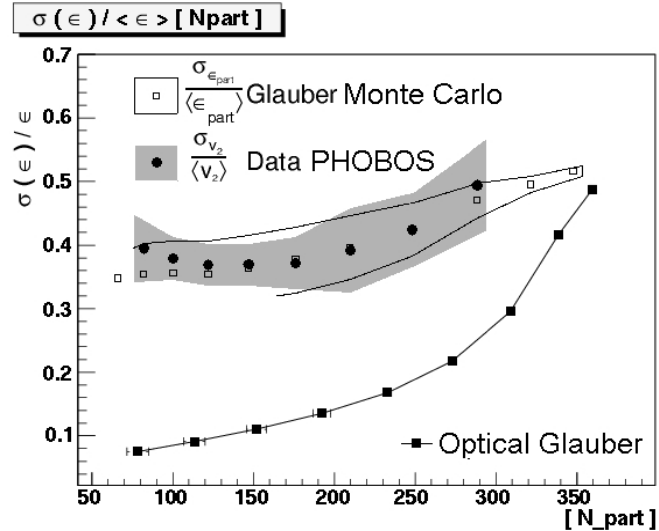}
\setlength{\captionmargin}{66pt}
\caption{\label{fig:FF5}
{\small
Relative fluctuations of eccentricity 
$\sigma(\varepsilon) / \langle \varepsilon\rangle$
obtained in optical Glauber model ($\beta_2^{\text{Au}} = -0.13$)
in comparison with Glauber Monte Carlo simulation \cite{PHOBOS}
(assuming spherical Au$^{197}$ nucleus) and relative elliptic 
flow fluctuations $\sigma (v_2)/\langle v_2 \rangle$
as measured by PHOBOS collaboration \cite{PHOBOS}. 
} }
\end{center}
\end{figure}

\vspace{-0.3cm}

We observe that elliptic flow fluctuations in the most central Au+Au collisions
observed by PHOBOS collaboration can be almost fully accounted for 
assuming Au$^{197}$ nucleus to have oblate ground-state deformation 
$|\beta_2| \approx 0.13$. 
Significantly lower strength of the elliptic flow fluctuations 
at smaller centralities predicted in our optical Glauber model
for Au+Au collisions 
is a simple consequence of neglecting the eccentricity fluctuations 
originating from variations of the individual positions of interacting nucleons
in the optical Glauber model.    

\section{Conclusions}
\indent

Based on the results obtained with our simple optical Glauber model
simulation we suggest to re-investigate consequeces of the possible
ground-state nuclear deformation of Au$^{197}$ nucleus
in Monte-Carlo Glauber model simulations. This would allow us
to clarify the influence of the predicted oblate deformation of 
Au$^{197}$ nucleus on interpretations of relativistic Au+Au 
heavy ion collision experiments.
We suggest to pay attention also to the
most-central Cu+Cu collisions at RHIC since both stable Cu isotopes 
are predicted to be deformed \cite{Moller}.

We have predicted that 
precise measurement of the elliptic flow and elliptic-flow fluctuations
in most central nucleus-nucleus collisions may allow one to determine
ground-state nuclear deformation of these nuclei. This should apply
also to nuclei with zero spin (e.g. Si$^{28}$, W$^{180}$) 
and unstable/exotic isotopes. 

\subsection*{Acknowledgement}
\indent

The author is grateful to the organizers of the 16th Conference of
Czech and Slovak Physicists for the kind hospitality in the
beautiful town of Hradec Kr\'alov\'e. 
This work has been supported by the Slovak Grant Agency 
under grant: N.2/7116/27.


\end{document}